\documentclass[]{llncs}
\usepackage{booktabs}
\usepackage{graphicx}
\usepackage{cite}
\usepackage[justification=centering, labelfont=bf]{caption}
\begin{document}
\title{DNS Tunneling: A Deep Learning based Lexicographical Detection Approach }

\author{Franco Palau\inst{1} \and
Carlos Catania\inst{1,3} \and
Jorge Guerra\inst{1} \and
Sebastian Garcia \inst{2} \and Maria Rigaki\inst{2}}

\institute{Universidad Nacional de Cuyo. Facultad de Ingenieria. LABSIN. Mendoza. Argentina\\
\email{franco.palau@ingenieria.uncuyo.edu.ar|harpo@ingenieria.uncuyo.edu.ar|\\jr@uncuyo.edu.ar} \and
CTU - Czech Technical University. Prague. Czech Republic
\email{sebastian.garcia@agents.fel.cvut.cz} \and
UCH - Universidad Champagnat. Mendoza. Argentina
}
\maketitle              % typeset the header of the contribution
\begin{abstract}
Domain Name Service is a trusted protocol made for name resolution, but during past years some approaches have been developed to use it for data transfer. DNS Tunneling is a method where data is encoded inside DNS queries, allowing information exchange through the DNS. This characteristic is attractive to hackers who exploit DNS Tunneling method to establish bidirectional communication with machines infected with malware with the objective of exfiltrating data or sending instructions in an obfuscated way. To detect these threats fast and accurately, the present work proposes a detection approach based on a Convolutional Neural Network (CNN) with a minimal architecture complexity. Due to the lack of quality datasets for evaluating DNS Tunneling connections, we also present a detailed construction and description of a novel dataset that contains DNS Tunneling domains generated with five well-known DNS tools. Despite its simple architecture, the resulting CNN model correctly detected more than 92\% of total Tunneling domains with a false positive rate close to 0.8\%.

\end{abstract}
\section{Introduction}
The Domain Name System (DNS) is a well-established and trusted protocol. It was originally made for name resolution and not for data transfer, so it is often not seen as a malicious communications and data exfiltration threat and therefore, organizations rarely analyze DNS packets for malicious activity. This lack of attention has made it attractive to attackers, who exploit the DNS protocol by using a technique called \textbf{DNS Tunneling}. With this method, data is encoded inside DNS queries and responses, allowing attackers to exchange information in an obfuscated way. 

A requirement to use DNS Tunneling is that the attacker must control a domain and a DNS server, which receives the DNS requests for the domain and act as an authoritative server for that domain in order to run the server-side tunneling and decoding programs. By doing so, the attacker can observe all incoming DNS queries and have control under the answers to the queries.

Consider a machine that has been comprised, i.e. infected with malware. DNS Tunneling can be used by the attacker in two ways \cite{das}:
\begin{itemize}
    \item to exfiltrate data from the compromised machine to an external server controlled by the attacker. 
    \item to establish a tunnel and send communications and instructions from the external server to the compromised machine.
\end{itemize}
If the attacker registers the domain \texttt{evilsite.com}, then data from comprised machine can be transmitted as a DNS request to\\ \texttt{\textless base64\_encoded\_data\textgreater.evilsite.com}. Then the server controlled by the attacker could send any command/data by responding to the query with a CNAME, TXT or NULL record as response. Hence, the attacker can receive and send data, i.e. a bidirectional data transfer channel is achieved using DNS tunneling. The advantages of this method include that DNS is almost always accessible in the network , no direct connection is established between the comprised machine and attacker, and pure data exfiltration (upstream only) is difficult to detect.

Catching threats which exploit the DNS has become a central topic in network security and various types of malicious domains detection methods have been proposed.
%Such methods can be categorized into two main classes \cite{sammour}. The first one, Traffic Analysis, where some significant features are identified from the traffic such as volume of DNS traffic, number of hostnames per domain, location and domain history. The second one, Payload Analysis, where the payload of a single request is analyzed to identify some features. As state in \cite{sans}, for payload analysis detection techniques, research on Domain Generation Algorithms (DGA) can be used for DNS Tunneling detection since DGA generated domains are abnormal in a similar way to names from data encoding.
A common approach for detecting these DNS Threats are the so-called lexicographical approaches. Under such approaches domains are classified by studying the statistical properties of the characters conforming the domain name, such as  the frequency distribution of properties, domain character length, Shannon entropy, the  vowel/consonant ratio and  dictionary-based similarity among others \cite{Antonakakis2012,das,bigram}. 

Several other DNS Threats detection approaches extended the idea of using the information provided by domain names properties to train a machine learning classifier such as Support Vector Machine (SVM), Naive Bayes (NB), Decision Tree (DT) \cite{sammour} or Logistic Regression \cite{das}. Recently, to avoid the need of designing the right set of features for training machine learning classifiers, some authors explored the application of Deep Learning (DL) techniques. In \cite{woodbridge}, an LSTM network was presented as a Domain Generation Algorithms (DGA) classifier reaching a 90\% detection rate. With a similar focus, in \cite{cacic} a DGA classifier was built, but in this case based on a simple 1D Convolutional neural network (1D-CNN) capable of detecting more than 97\% of total DGA domains with a false positive rate close to 0.7\%.

As state in \cite{sans}, research on DGA detection can be used for DNS Tunneling detection since DGA generated domains are abnormal in a similar way to names from data encoding, so it is reasonable to apply the same detection approach. Following this idea, in the present work we focus on analyzing the DNS Tunneling detection performance of a 1D-CNN with an architecture similar to \cite{cacic}. In addition and given the lack of quality datasets for evaluating  DNS tunneling connections, we create a new dataset from five well-known DNS tunneling tools. The dataset was created on Virtual Machine infrastructure using a time injection methodology\cite{icissp18} for executing and logging several DNS tunneling tools.

The main contributions of the present article are:
\begin{itemize}
    \item A 1D convolutional neural network (1D-CNN) capable of detecting DNS Tunneling in Domain names.
    \item The evaluation of the 1D-CNN model on a new generated dataset following a well defined methodology for labeling tunneling connections.
\end{itemize}

The rest of this paper is organized as follows: Section \ref{sec:The Neural Network Model} describes the network architecture, Section \ref{sec: Dataset} explains the dataset construction, Section \ref{sec:exp-design} illustrates the experimental design while Section \ref{sec:results} details the results, and finally in Section \ref{sec:conclusions}, we present the conclusions. 

\section{Deep Learning model Lexicographical  Detection}
\label{sec:The Neural Network Model}

The Neural Network Architecture model is a 1D Convolutional Neural Network (CNN). This CNN is composed of three main layers. The first one is an \textit{Embedding} layer, then there is a \textit{Conv1D} layer, and finally a Dense fully connected layer. The first two layers are the most relevant components of the architecture regarding the problem of detecting Tunneling domains. Both layers are responsible for learning the feature representation in order to feed the third Dense and fully connected layer. Beside the three layers previously described, the complete Neural Network Architecture includes some other layers for dealing with the dimensions output of the \textit{Conv1D} layer as well as layers for representing the input domain and the output probability. A detail of the complete architecture together with the used activation functions is shown in Table \ref{tab:complete-nn-arch}, whereas the three main layers are described in the following subsections. 
\begin{table}[]
\centering
\caption{The complete network architecture including the corresponding output dimensions and activation function used in each layer}
\label{tab:complete-nn-arch}
\begin{tabular}{@{}ll@{}}
\toprule
Layer (type)                    & Activation Function \\ \midrule
input (Input Layer)              & -                \\
\textbf{embedding (Embedding)}  & -                \\
\textbf{conv1d (Conv1D)}        & relu             \\
%flatten (Flatten)               & -                \\
\textbf{dense\_1 (Dense})       & relu             \\
dense\_2 (Dense)                & sigmoid           \\ \bottomrule
\end{tabular}
\end{table}

\subsection{Embedding Layer}

A character embedding consists in projecting $l$-length sequences of input characters into a sequence of vectors $R^{lxd}$, where $l$ has to be determined from the information provided by the sequences in the training set and $d$ is a free parameter indicating the dimension of the resulting matrix~\cite{woodbridge}. By using an \textit{Embedding} layer in the architecture, the neural network  learns in an efficient manner the optimal set of features that represent the input data. 

\subsection{1D Convolutional Layer}

\begin{figure}[h!]
	\centering
    \includegraphics[width=0.7\columnwidth, trim={0 0 0 0},clip]{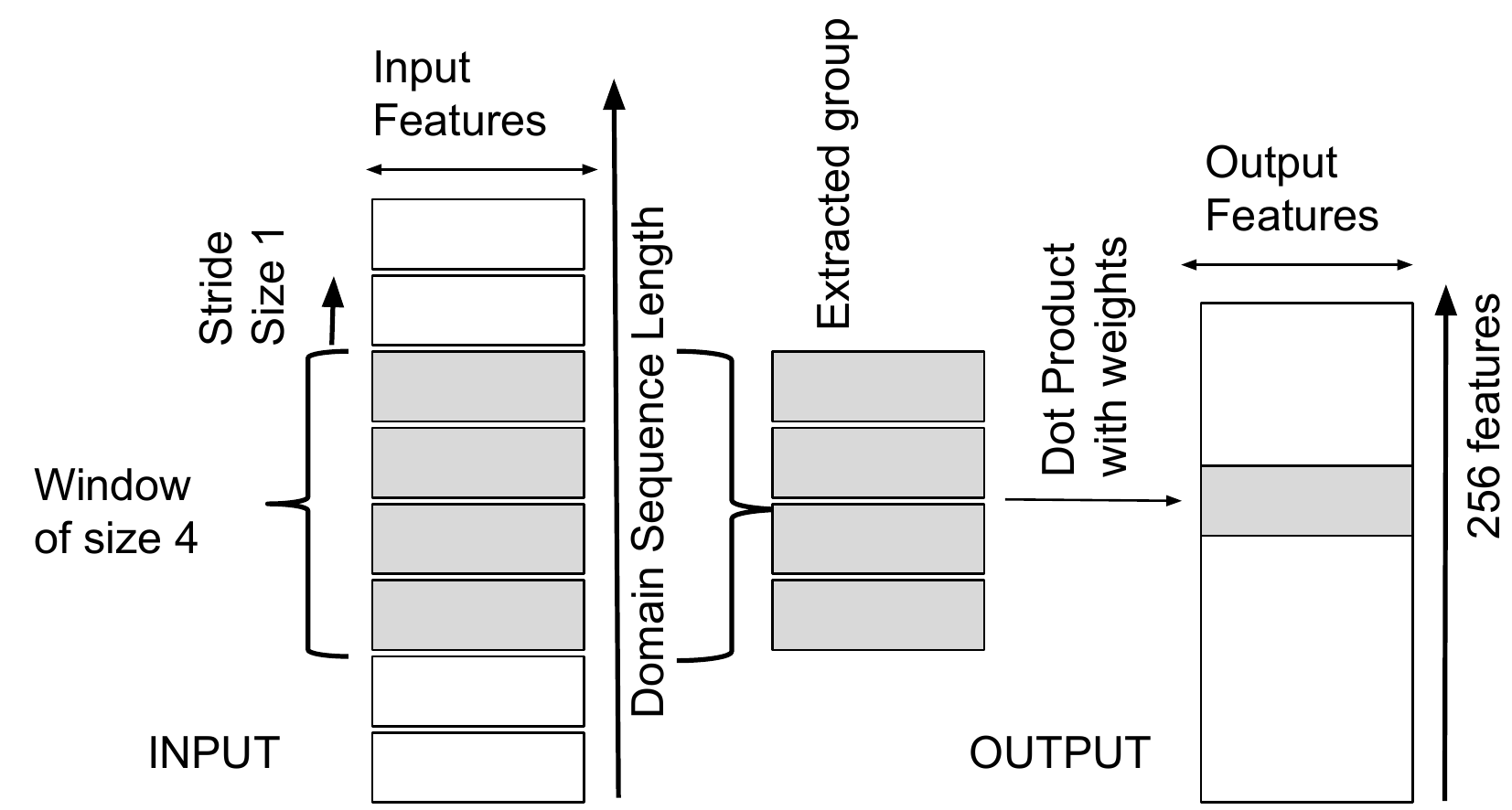}     
    \caption{\label{fig:conv-work} {Feature extraction process of the 1D convolutional layer}}
\end{figure}

The \textit{Conv1D} layer refers to a convolutional network layer over one dimension. For the Tunneling detection problem, such dimension consists of the length of the domain name sequence. The convolutional layer is composed of a set of convolutional filters that are applied to different portions of the domain name. A visual example of the feature extraction process for a 1D Conv layer is shown in Fig. ~\ref{fig:conv-work}.
The figure depicts a 1D convolutional layer constructing 256 filters (features) ($nf=256$), with a window (kernel) size of 4 ($ks=4$) and a stride length value of 1 ($sl=1$). The layer selects from groups (also referred as patches) of 4 characters to apply the convolutional filters, and continues shifting one character at a time (stride value) applying the same convolutions filter over the rest of the sequence. Consequently, the neural network generates 4-grams features. These features represent the discriminative power of these group of letters in the domain names.

By applying the same filter all over the sequence the required computation time is considerable reduced when compared with traditional Multilayer Perceptron layers. Additionally,  since a  convolutional kernel independently  operates on each 4-gram it is possible to go over the entire input layer concurrently. This paralellization and its consequent low computing time is one of the major benefits of using convolutional networks instead of other deep learning approaches usually used for text processing such as Long Short Term Memory (LSTM) \cite{hochreiter97,torres2016, woodbridge}.

\subsection{Dense Layers}  
The features extracted by both  previous layers are be used by traditional Multilayer Perceptron Network (MLP) to output the probability of a given domain belongs to Tunneling or Normal class. MLP is composed of two layers: A first fully connected layer of size $hn$  (\textit{Dense} layer ) connected to a second \textit{Dense} layer of size $1$  used for actually giving the probability output about the considered domain.

\section{Dataset}
\label{sec: Dataset}
\subsection{A Methodology for Collecting Tunneling DNS Data}
\label{sec: collecting_tun}
\begin{figure}[h!]
	\centering
    \includegraphics[width=1\columnwidth, trim={0 0 0 0}]{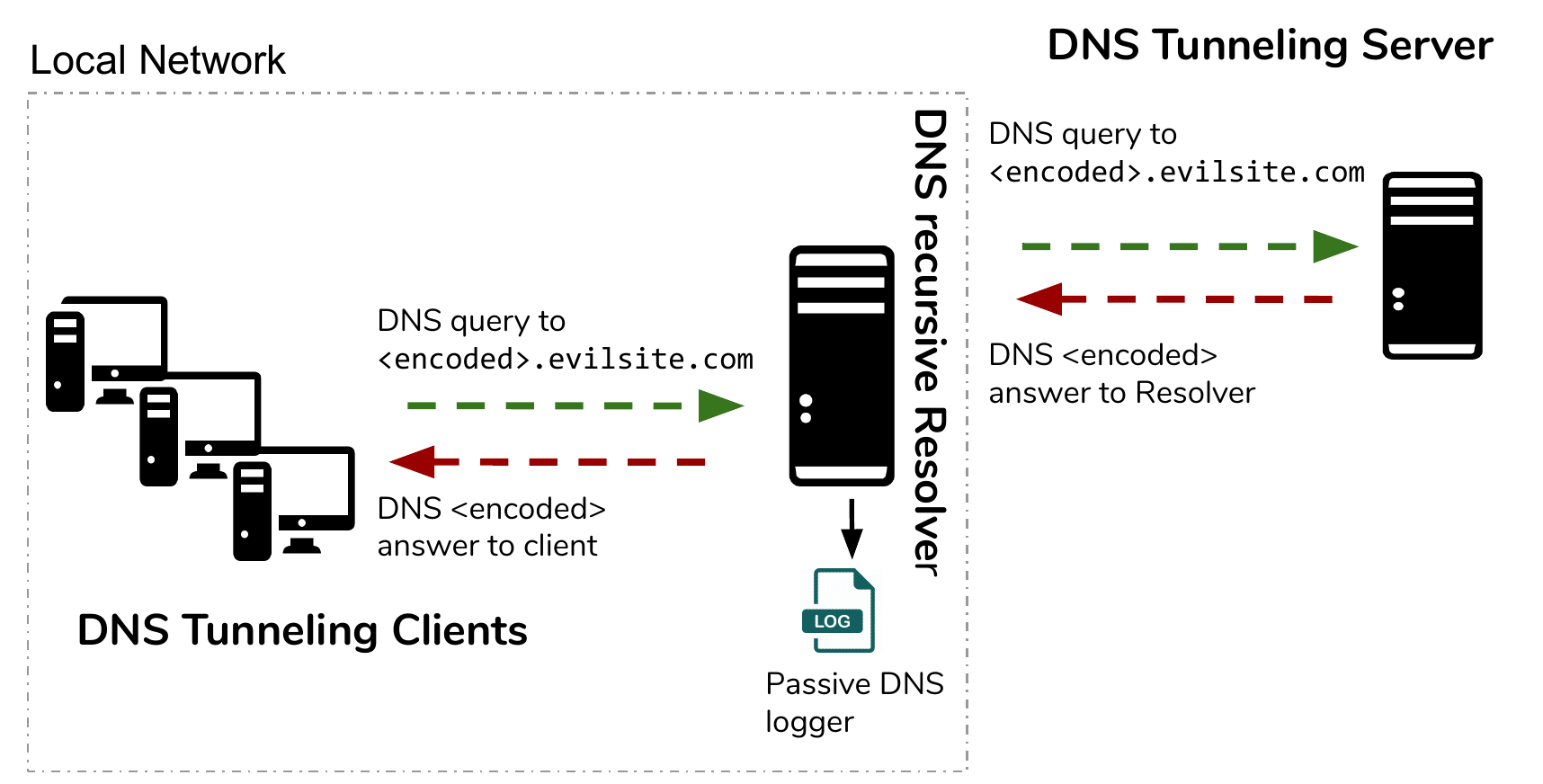}    
    \caption{\label{fig:vm-platform} {Virtual Machine infrastructure (VMI) for acquiring DNS tunneling data}}
\end{figure}

A Virtual Machine Infrastructure (VMI) including all the required components for performing a DNS tunneling connection was deployed in order to proceed with the  acquisition of DNS tunneling data.  In the diagram from Fig. \ref{fig:vm-platform} it is possible to recognize the three main components involved in DNS tunneling: (a) the clients, which are a group a (possible) compromised machines inside the local network, (b) a local DNS server in charge of logging and resolving all local clients DNS requests.  and (c) a computer outside the local network where DNS tunneling server side is running. 
In addition, two domain names were registered in order to provide authoritative answers: \texttt{harpozedcompute.com} and \texttt{securitytesting.online}. Both domains were properly configured for pointing to computer running the DNS tunneling servers.
% ,cliphttps://www.overleaf.c qom/project/5d7788b467fd890001b8f250

The use of VMI facilitates the execution of different DNS tunneling tools under a controlled environment. In this work, five different DNS tunneling tools were executed inside the VMI. The name along with a summary description of each DNS tools is shown in Table \ref{tab:dnstools}. Four out the five tools are oriented to high  throughput connections. The only exception is the \texttt{DNSexfiltrator} tool which performs sporadic data exchanges using DNS. Certainly, the detection of such low throughput data exchanges results in a considerable more difficult task.

\begin{table}[h]
\centering
\caption{DNS tunneling tools executed inside the VMI}
\label{tab:dnstools}
\begin{tabular}{@{}lll@{}}
\toprule
\textbf{Tool} & \textbf{Description} & \textbf{Throughput} \\ \midrule
\textbf{\texttt{tuns}} & IpV4 over DNS tunneling & High \\
\textbf{\texttt{dnscat2}} & C\&C oriented DNS tunneling & High \\
\textbf{\texttt{dns2tcp}} & TCP over DNS tunneling & High \\
\textbf{\texttt{iodine}} & IpV4 over DNS tunneling & High \\
\textbf{\texttt{DNSexfiltrator}} & simple data exchange over DNS & Low \\ \bottomrule
\end{tabular}
\end{table}

A set of Unix shell scripts were implemented for automatizing different common actions performed through the tunnel. In the case of the high throughput tools, the script consisted of a number of ICMP ECHO REQUEST packets and the transfer of files of different sizes. In the case of \texttt{DNSExfiltrator}, only the transfer were implemented. 

A simple approach for collecting all the tunneling conections inside the recursive DNS logs consists of searching for all requests coming from \texttt{harpozedcomput.com} and \texttt{securitytesting.online}. In addition, for precisely recognizing the request corresponding to each tool, a time injection strategy\cite{icissp18} was used. Under such strategy, the script execution timestamp is recorded for further timestamp matching with the recursive DNS server logs in the VMI. 

Finally, since both \texttt{tuns} and \texttt{dns2tcp} failed at establishing the tunnel connections, neither file transfer nor ICMP packet were transmitted. However, the repeated failed attempts were still logged.

\subsection{Dataset Description}
The dataset used in this paper contains both DNS Tunneling and normal domain names.The resulting distribution for both classes can be observed in Table \ref{tab:dnstools-distribution}.
\begin{table}[h]
\centering
\caption{DNS tunneling tools executed inside the VMI}
\label{tab:dnstools-distribution}
\begin{tabular}{@{}lll@{}}
\toprule
\textbf{Class} & \textbf{Source} & \textbf{Total} \\ \midrule
Tunneling & \textbf{\texttt{dnscat2}} & 23 \\
 & \textbf{\texttt{DNSexfiltrator}} & 78 \\ 
 & \textbf{\texttt{iodine}} & 346  \\
 &\textbf{\texttt{Not specified}} & 7553 \\ 
\midrule
& \textbf{\texttt{cz}} & 511 \\
Normal & \textbf{\texttt{bambenek}} & 3000 \\ 
 & \textbf{\texttt{alexa}} & 5000 \\ 
  \bottomrule
\end{tabular}
\end{table}

For DNS Tunneling, 8000 tunnel domains were generated as explained in Section \ref{sec: collecting_tun}. Three tunneling tools were discriminated,\texttt{dnscat2} \cite{dnscat2}, \texttt{iodine} \cite{iodine} and  \texttt{DNSExfiltrator} \cite{dnsExfiltrator}. Several DNS requests coming from unspecified tunneling tools,  as well as failed attempts to establish the tunneling connection, are refered as \texttt{Not Specified} in the table.

In order to balance the dataset for both  domains classes (i.e. DNS Tunneling and normal), we collected 8000 normal domains that come from the Alexa top one million domains and the Bambenek Consulting feed. Finally, a small portion of the dataset contain commonly accessed domains inside Czech Republic domain name servers. The Czech language is particularly difficult to detect given its low vocal/consonant ratio.

The complete dataset detailed in this section was split into two new datasets: the first one, containing 80\% of total entries, is used for the training and tuning of the network. The second one, containing the remaining 20\% of total entries, is used for the evaluation of the proposed model on unseen domains. Both training and testing datasets can be found in \cite{dataset}.

\section{Experiment Design}
\label{sec:exp-design}
In this section, we describe the selected metrics and the hyper-parameters fine-tuning methodology for improving the model results.

\subsection{Metrics}
\label{sec: metrics}
Several standard metrics are used for evaluating the network. These metrics are \textbf{Precision}, \textbf{Recall} or True Positive Rate (\textbf{TPR}), False Positive Rate (\textbf{FPR}) and \textbf{F1-Score}. Precision is computed as the ratio of items correctly identified as positive out of total items identified as positive while Recall is computed as the ratio of items correctly identified as positive out of total true positives. The F1-Score is the harmonic mean of Precision and Recall. Finally,  False Positive Rate is computed as the ratio between items incorrectly identified as positive and the total number of actual negative predictions. 

\subsection{Hyper Parameters Tuning}
The proposed model possesses many hyper-parameters that need to be tuned. In
order to complete the fine tuning task, a traditional Grid Search was conducted.
Among all the possible hyperparameters, we particularly focused on finding the optimal values related to the Embedding, Conv1D and Dense layers. 
For a robust estimation, the evaluation of each parameter combination was carried out using a k-fold cross validation with k = 5 folds. The 1D-CNN was trained using the back propagation algorithm \cite{backpropagation} considering the Adaptive Moment Estimation optimizer \cite{adam}. The 1D-CNN training was carried out during 10 epochs. Table \ref{tab:tuning} shows the parameter combinations with the best performance detection in terms of the F1-Score.

\begin{table}[h]
\centering
\caption{Best Hyper-parameters subset. For space reason only the higher average F1-Score parameter combination is shown. This combination was chosen for all the remaining experiments.}
\label{tab:tuning}
\resizebox{0.8\textwidth}{!}{%
\begin{tabular}{@{}ccccccccc@{}}
\toprule
\textbf{avg. F1}  & \textbf{sd}       & \textit{nf}   & \textit{ks} & \textit{sl} & \textit{d}   & \textit{l}  & \textit{hn}  & \textbf{parameters} \\ \midrule
\textbf{0.936820} & \textbf{0.011219} & \textbf{1024} & \textbf{4}  & \textbf{1}  & \textbf{100} & \textbf{45} & \textbf{256} & 11,425,685               \\ \bottomrule
\end{tabular}%
}
\end{table}

\section{Results}
\label{sec:results}
The hyper-parameters selected in the previous section were employed for training the network. Then, the model was tested using the explained testing dataset. The experiment results are shown in Table \ref{tab:metrics} where metrics illustrated in Section \ref{sec: metrics} are calculated with a decision boundary threshold set to 0.90. The support (size of test set) is given in the last column.
%Figure \ref{fig:conf_matrix} with the resulting Confusion Matrix. Based on this,  and described in Table \ref{tab:metrics}. 

%\begin{figure}[]
%\centering
%\includegraphics[width=0.7\textwidth]{conf_matrix.png}
%\caption{Confusion matrix}
%\label{fig:conf_matrix}
%\end{figure}

The resulting F1-Score around 96\% and the low FPR achieved for the two considered classes (i.e Normal and DNS Tunneling), demonstrate that the 1D-CNN detection method was capable of extracting common patterns that are important to discriminating malicious domains (DNS Tunneling) from non-malicious domains. 

\begin{table}[]
\centering
\caption{Resulting metrics of the proposed model evaluation on unseen domains}
\label{tab:metrics}
\resizebox{\textwidth}{!}{%
\begin{tabular}{@{}llclclclclc@{}}
\toprule
\multicolumn{1}{c}{\textbf{Domain Type}} &  & \textbf{Precision} &  & \textbf{Recall (TPR)} &  & \textbf{FPR} &  & \textbf{F1-Score} &  & \textbf{Support} \\ \midrule
\textit{Normal}                          &  & 0.9342                  &  & 0.9948                     &  & 0.0762            &  & 0.9635                 &  & 1726             \\
\textit{Tunneling}                       &  & 0.9939                  &  & 0.9237                     &  & 0.0052            &  & 0.9575                 &  & 1586             \\ \bottomrule
\end{tabular}%
}
\end{table}

\begin{figure}[h]
\centering
\includegraphics[width=1\textwidth]{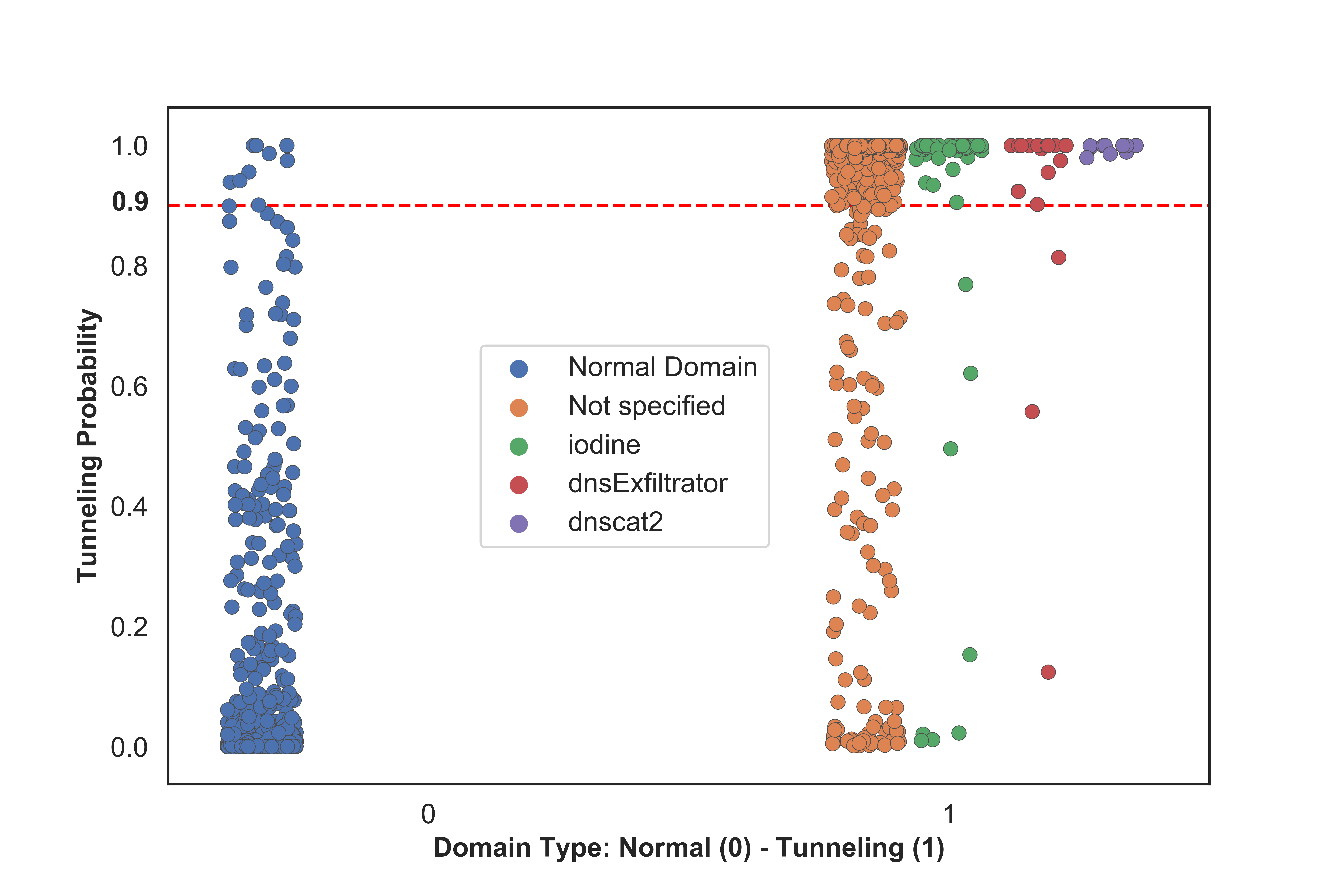}
\caption{Tunneling probability explicit for each domain in testing set. Dots corresponding to DNS Tunneling class have been colorized according to their tunneling tool}
\label{fig:tool}
\end{figure}

In Figure \ref{fig:tool} we show the probability of being Tunneling given by the network to each domain in testing set. Dots located on the left correspond to the true Normal domains and dots located on the right are the true Tunneling domains. The decision boundary threshold of 0.90 is plotted as a red line which indicates that all dots above this line are classified as Tunneling domains whereas dots below are classified as Normal domains.

When Tunneling domains are discriminated by the tunneling tool used to generated them, it can be seen that we were able to identify all Tunneling domains generated with \texttt{dnscat2} (violet dots), while in the case of \texttt{DNSExfiltrator} (red dots) and \texttt{iodine} (green dots), we detected correctly 85\% and 87\% of these Tunneling domains respectively. 
There is a group of domains (orange dots) that, as explained in Section \ref{sec: Dataset}, failed at establishing the tunnel connection, so they don't have a tunneling tool specified. We detected correctly 96\% of these domains, being these results consistent with the overall performance of the network.

\newpage
\section{Concluding Remarks}
\label{sec:conclusions}
In the present work, we explored the viability of 1D-CNN for lexicographical DNS Tunneling detection. 

A new dataset containing 8000 DNS requests was specially created for evaluating  the proposed model on  domains generated with 5 common tunneling tools as well as normal domains.

The dataset was properly split into training and testing sets. A hyper-parameters grid search was conducted on training set and the best resulting model was then evaluated on the testing set.

The resulting 1D-CNN was able to detect 99\% of total normal domains and 92\% of total Tunneling domains, with a FPR of 0.07\% and 0.005\% respectively. Although its simple architecture, such
results make it suitable for real-life networks. 

Inspecting the Tunneling tool that were used to generate each malicious domain, we observed that we were able to identify all domains generated with \texttt{dnscat2} and around 86\% for \texttt{DNSExfiltrator} and \texttt{iodine}. An analysis of the lexicographical similarities and differences between domains generated with these tools is left for future work. Moreover, since both \texttt{tuns} and \texttt{dns2tcp} failed at establishing the tunnel connections, domains considered for the analysis of detection rate discriminated by tunneling tool  correspond to only 3 different tools. Therefore, it could be necessary an in-depth analysis of how including more tunneling tools affects the performance of the network.

\section*{Acknowledgments}
The authors would like to thank the financial support received by Universidad Champagnat, CVUT and UNCuyo during this work. In particular the founding provided by the Czech TACR project no. TH02010990 and the PICT 2015-1435 granted by ANPCyT. The authors would also like to specially thank Whalebone s.r.o., whose technical support and help have been fundamental to the complete research process. In addition, we want to gratefully acknowledge the support of NVIDIA Corporation with the donation of the Titan V GPU used for this research.

% ---- Bibliography ----

\bibliographystyle{splncs04}
\bibliography{deeptunnel.bib}
\end{document}